\documentclass[journal]{IEEEtran}
\usepackage{cite}
\usepackage{amsmath,amssymb,amsfonts}
\usepackage{algorithmic}
\usepackage{graphicx}
\usepackage{textcomp}
\usepackage{xcolor}
\usepackage{booktabs}
\usepackage{epstopdf}
\usepackage{footnote}
\usepackage{amsmath}
\usepackage{algorithmic}
\usepackage{multicol}
\usepackage[caption=false,font=footnotesize]{subfig}
\usepackage{mathtools}
\usepackage{amsfonts}
\usepackage{array}
\usepackage{diagbox}
\usepackage{stfloats}
\usepackage{url}
\usepackage{verbatim}
\usepackage{graphicx}
\usepackage{balance}

\def\BibTeX{{\rm B\kern-.05em{\sc i\kern-.025em b}\kern-.08em
    T\kern-.1667em\lower.7ex\hbox{E}\kern-.125emX}}

\begin{document}
\title{Cross Far- and Near-field Wireless Communications in Terahertz Ultra-large Antenna Array Systems
}
\author{\IEEEauthorblockN{
    Chong~Han,~Yuhang~Chen, Longfei~Yan, Zhi~Chen, and Linglong~Dai
	}
		\thanks{
			Chong Han is with Terahertz Wireless Communications (TWC) Laboratory,
			and with Department of Electronic Engineering and Cooperative Medianet
			Innovation Center (CMIC), Shanghai Jiao Tong University, Shanghai 200240, China (e-mail:
			chong.han@sjtu.edu.cn).

			Yuhang Chen and Longfei Yan are with Terahertz Wireless Communications (TWC) Laboratory, Shanghai Jiao Tong University, Shanghai 200240, China (e-mails: \{yuhang.chen, longfei.yan\}@sjtu.edu.cn).
			
			Zhi Chen is with the National Key Laboratory of Science and Technology	Communications, University of Electronic Science and Technology of China,
			Chengdu 611730, China (e-mail: chenzhi@uestc.edu.cn).
			
			Linglong Dai is with the Beijing National Research Center for Information Science and Technology (BNRist) as well as the Department of Electronic Engineering, Tsinghua University, Beijing 100084, China 
			(e-mail: daill@tsinghua.edu.cn).
			}}

\maketitle

\begin{abstract}\label{sec_abstract}
	Terahertz (THz) band owning the abundant multi-ten-GHz bandwidth is capable to support Terabit-per-second wireless communications, which is a pillar technology for 6G and beyond systems. 
	With sub-millimeter-long antennas, ultra-massive (UM)-MIMO and reflective intelligent surface (RIS) systems with thousands of array elements are exploited to effectively combat the distance limitation and blockage problems, which compose a promising THz ultra-large antenna array (ULAA) system. 
	As a combined effect of wavelength and array aperture, the resulting coverage of THz systems ranges from near-field to far-field, leading to a new paradigm of cross-field communications. 
	Although channel models, communications theories, and networking strategies have been studied for far-field and near-field separately, the unified design of cross-field communications that achieve high spectral efficiency and low complexity is still missing.
	In this article, the challenges and features of THz ULAA cross-field communications are investigated.
	Furthermore, cross-field solutions in three perspectives are presented, 
	including a hybrid spherical- and planar-wave channel model, cross-field channel estimation, and widely-spaced multi-subarray hybrid beamforming, where a subarray as a basic unit in THz ULAA systems is exploited.
	The approximation error of channel modeling accuracy, spectral efficiency, and estimation error of these designs are numerically evaluated.
	Finally, as a roadmap of THz ULAA cross-field communications, multiple open problems and potential research directions are elaborated. 
\end{abstract}

\section{Introduction}
\label{sec_introduction}
The boosting wireless traffic demands inspire increasing research attention to the exploration of future six-generation (6G) and beyond communication systems.
To break the bottleneck of limited bandwidth, 
the new spectrum in the Terahertz (THz) band, which ranges from 0.1 to 10~THz and possesses abundant continuous bandwidth over tens of GHz, is regarded as a pillar technology to meet the terabits per second (Tbps) data rates for 6G communication systems~\cite{ref_TSR_THz}.
On the downside, as one major challenge, the THz wave suffers from severe propagation losses caused by large free space attenuation and strong molecular absorption, which thus limit the communication distance~\cite{ref_THz_2030b}.
Moreover, being reliable on the line-of-sight (LoS) transmission due to diffuse scattering, 
signal propagation in the THz band is prone to be blocked by various obstacles in the communication channel.

Fortunately, the sub-millimeter wavelength in the THz band makes it possible to compactly arrange hundreds and even thousands of antennas in a small area, which composes ultra-massive multi-input multi-output (UM-MIMO) systems~\cite{ref_25_Heath}.
By generating razor-sharp narrow beams with high beamforming gain, UM-MIMO systems can effectively combat the distance limitation problem. 
Furthermore, reflective intelligent surface (RIS) systems, by arranging a bountiful number of passive reflecting elements, can effectively solve the blockage problem~\cite{ref_THz_2030b}.
By employing UM-MIMO and RIS, the composed THz ultra-large antenna array (ULAA) system operates with a massive number of sub-millimeter-long active and passive elements. 

In THz ULAA systems, one question arises: do THz signals propagate in \textit{far-field}, \textit{near-field}, or \textit{cross-field} of the antenna array?
As a classic boundary between far- and near-field, the Rayleigh distance for the antenna array is proportional to the square of the array aperture divided by the wavelength, which is calculated as $\frac{2S^2}{\lambda}$, where $S$ denotes the array aperture, and $\lambda$ represents the wavelength~\cite{ref_nearfield_dai}.
As a result, the Rayleigh distance grows with the rapid increment of array size, as well as the decrement of wavelength.
Considering a uniform planar array (UPA) with 1024 elements at the transmitter (Tx) and receiver (Rx) operating at 0.3 THz with half-wavelength spacing, 
the Rayleigh distance is calculated as around 1m.
This distance could be further increased to tens and even hundreds of meters in THz widely-spaced multiple-subarrays (WSMS) systems~\cite{ref_hybrid_beamforming}, i.e., co-located yet distributed arrays.

\begin{figure*}[t]
	\centering
	{\includegraphics[width= 0.95\textwidth]{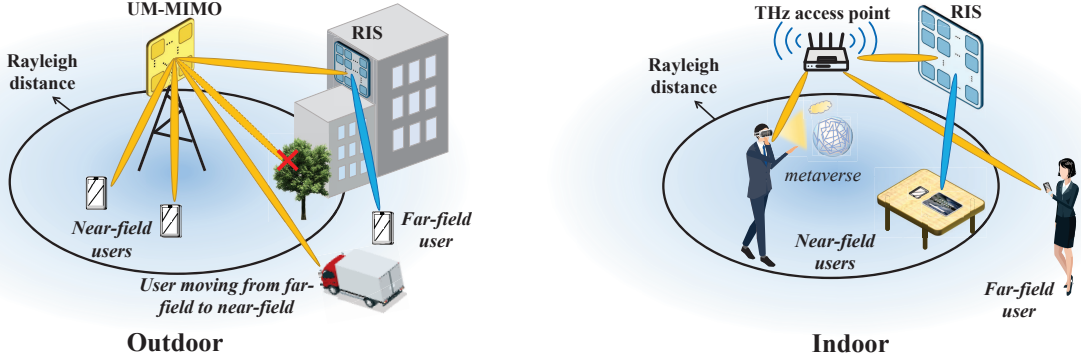}}
	\caption{Illustration of typical THz ULAA cross-field communications.} 
 \vspace{-5mm}
	\label{fig_ULAA_applications}
\end{figure*}

With such a system configuration, 
communications range from near-field to far-field in typical indoor and outdoor scenarios, e.g., metaverse, vehicular communications, etc. 
Therefore, a new paradigm of cross-field communications in THz ULAA systems is emerging, in typical outdoor and indoor scenarios of THz ULAA cross-field communications as shown in Fig.~\ref{fig_ULAA_applications}. 
It is worth noticing that the definition of cross-field is similar to \textit{hybrid-field} in traditional radar and communication systems~\cite{ref_Hybrid_CE_GC}.
By contrast, the hybrid-field stresses the coexistence of far- and near-field paths of the channel, while the cross-field emphasizes the transmission distance spans from near- to far-field.

Although UM-MIMO and RIS systems have been recently exploited, such as channel modeling and analysis~\cite{ref_SW_PW_Modeling, ref_powerscaling_bj}, channel estimation~\cite{ref_CE_ELMaMIMO,ref_OMP}, beam training~\cite{ref_subarray_lin, ref_near_training},  beamforming design~\cite{ref_25_Heath} and multiple access~\cite{ref_Dai_LDMA}, among other topics, the separate designs in either far-field or near-field are invalid as unified solutions of the cross-field communications. 
On one hand, the planar-wave model (PWM) where the wave propagation is approximated as a plane is adopted in the far-field. The omission of the non-linear phase term results in low complexity and satisfactory accuracy, which, however, causes inaccuracy and thus transmission performance degradation when it comes to the near-field. 
On the other hand, the spherical-wave model (SWM) derived from the electromagnetic (EM) wave theory is considered as the ground-truth. SWM is useful when the communication range is shorter than the Rayleigh distance, i.e., in the near-field.
Nevertheless, the high complexity makes SWM impractical, especially in the far-field.
In conclusion, 
neither SWM nor PWM can be efficiently applicable in THz ULAA cross-field communications. 

In this article, we investigate the challenges of THz ULAA cross-field communications in Sec.~\ref{sec_challenges} from channel and ULAA architecture perspectives.
The analysis indicates that in contrast to microwave multi-antenna systems, THz ULAA systems could take a subarray instead of an antenna as a unit, based on the practical consideration of hardware and transmission design.
The cross-field solutions are presented in three perspectives in Sec.~\ref{sec_hybrid_solutions}, including hybrid spherical- and planar-wave channel model (HSPM), compressive sensing (CS)-based cross-field channel estimation and WSMS hybrid beamforming, all of which are inspired by the subarray units in THz ULAA systems.
Numerical results are presented in Sec.~\ref{sec_performance_evaluation} to illustrate the modeling accuracy, estimation error and spectral efficiency of these methods, respectively.
Multiple open problems and research directions for 6G THz ULAA cross-field communications are elaborated in Sec.~\ref{sec_open_problems}. Finally, this paper is summarized in~Sec.~\ref{sec_conclustion}.

\section{Challenges of THz ULAA Cross-Field Communications}
\label{sec_challenges}
In this section, the challenges and unique features of THz ULAA cross-field communications from channel and ULAA architecture perspectives are elaborated.

\subsection{Challenges From Cross-field Channel Perspective}
\label{subsec_challenge_channel}
In terms of channel modeling, since THz ULAA communication distance crosses far- and near-field, either SWM or PWM models result in considerably high complexity or modeling accuracy degradation. 
On one hand, in SWM, considering one antenna pair between Tx and Rx for a propagation path, the amplitude and phase of the channel response are individually calculated, leading SWM the ground-truth channel model. 
However, the number of parameters to determine SWM is proportional to the production of the number of antennas and the number of propagation paths, which reaches $10^7$ if we consider 1024 antennas at Tx and Rx, and sparse THz multi-paths (e.g., less than 10)~\cite{ref_hybrid_beamforming}. 

On the other hand, as an approximation of SWM in the far-field region, PWM regards the radio-wave front as a plane.
The channel response of one antenna can be derived from the phase difference between it and the reference antenna. 
Therefore, to determine PWM, only channel parameters of the reference antenna are required, of which the number is comparable to the number of THz sparse multi-paths, e.g. 10, and thus much smaller than that of SWM.
Nevertheless, PWM becomes inaccurate in the near-field region due to phase approximation error by the planar-wave propagation assumption~\cite{ref_HSPM}.

The selection of transmission technologies is heavily influenced by channel models, therefore, most of the literature transmission technologies are specifically designed based on either PWM or SWM.
However, neither SWM nor PWM based technologies are effectively suitable for cross-field communications.
By deploying SWM, transmission designs have to consider the massive number of parameters in SWM, leading which suffer from high complexity. 
For example, angle and propagation distance information should be considered to determine the phases of SWM elements, which, therefore, should be considered in channel estimation and beam training.
For comparison, only angle information is considered in PWM-based designs. The additional distance domain information results in high complexity of channel estimation and beam training~\cite{ref_CE_ELMaMIMO,ref_near_training}.

\begin{figure*}[t]
	\setlength{\belowcaptionskip}{0pt}
	\centering
    \subfloat{
        \includegraphics[width= 0.4\textwidth]{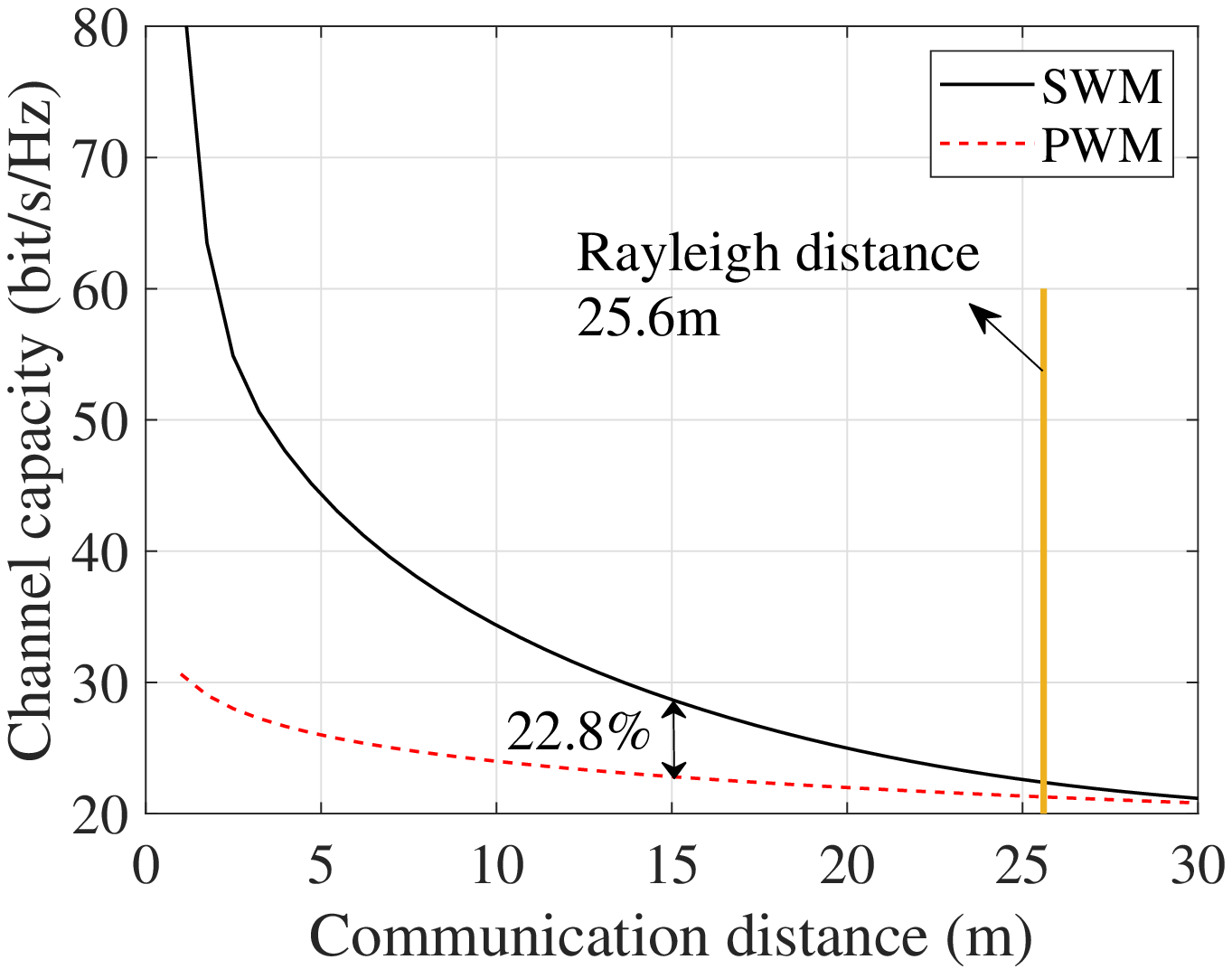} }
        \hspace{0.5mm} 
    \subfloat{
        \includegraphics[width= 0.4\textwidth]{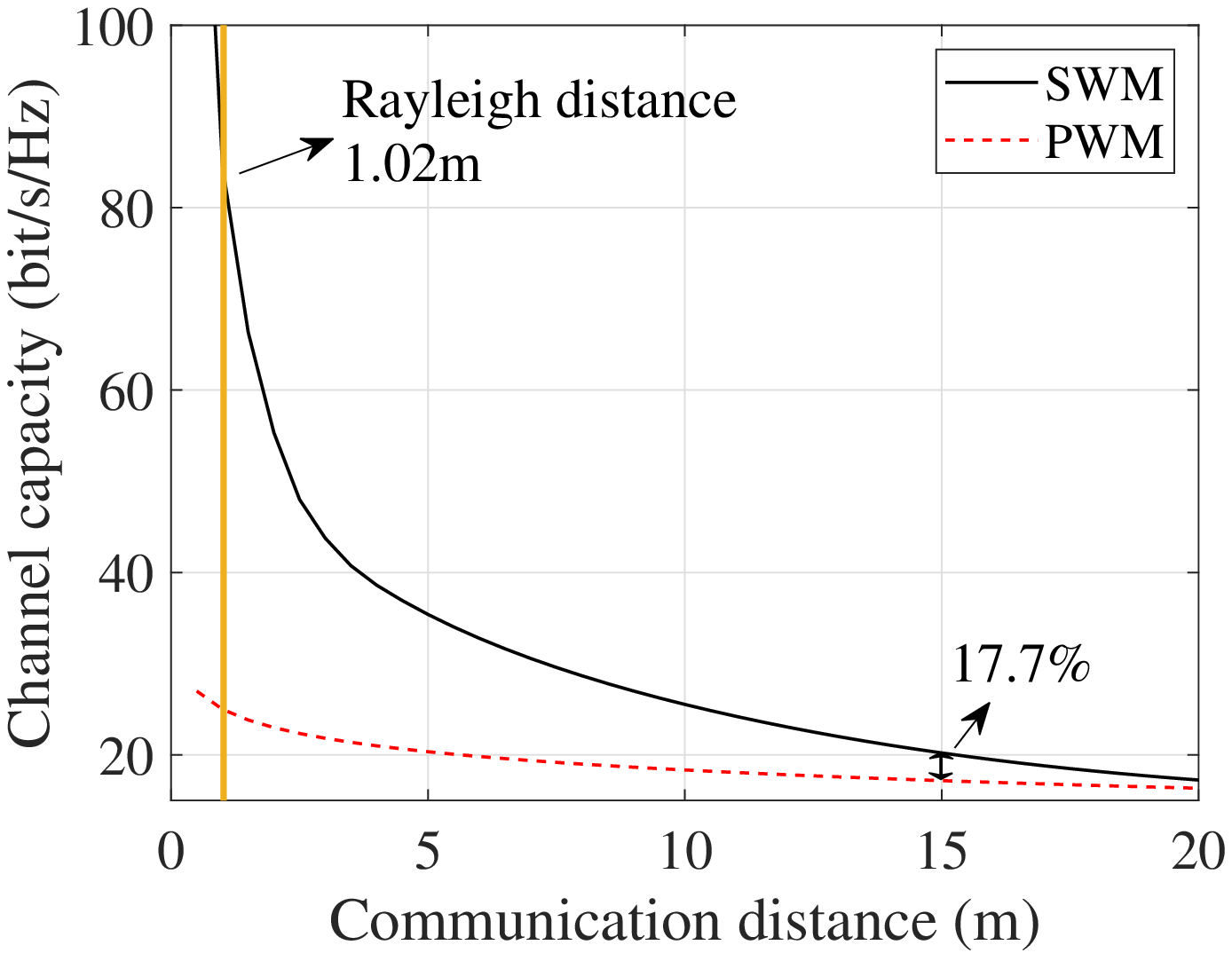} }
    \caption{{Channel capacity based on PWM and SWM in different communication distances and frequencies. 
    The WSMS with 1024 elements and 4 subarrays is considered in the left figure, the subarray spacing is 64 times the wavelength.
    The uniform planar array (UPA) with 1024 elements is deployed in the right figure. Communication frequency is fixed as 300 GHz.}}
    \label{fig_capacity_comp}
    \vspace{-5mm}
\end{figure*}

By contrast, due to the inaccuracy of PWM in the near-field, transmission technologies based on PWM, such as channel estimation and beam training, suffer from significant estimation and training accuracy losses, respectively, especially in the near-field region~\cite{ref_DSE_SSE}.
In addition, examples of channel capacity based on PWM and SWM are shown in Fig.~\ref{fig_capacity_comp}.
For systems equipped with 1024-element WSMS operating at 300~GHz, the capacity based on PWM is $22.8\%$ lower than that of SWM at a 15m communication distance. 
The capacity difference amplifies as the distance reduces to the near-field, suggesting the inaccuracy of PWM.
Moreover, it is worth noticing that there still exists a capacity gap between PWM and SWM even when the communication distance exceeds the Rayleigh distance.
This needs to be highlighted that the Rayleigh distance is derived based on the phase difference between PWM and SWM above a certain discrepancy. By viewing the mismatch of the channel capacity, one can tell Rayleigh distance itself is an approximate boundary between the far- and near-fields, instead of zero to one leap.

\subsection{Challenges From ULAA Architecture Perspective}
\label{subsec_challenge_architecture}
The design of THz ULAA architecture and transmission technologies should carefully consider the THz hardware complexity, power consumption, and propagation features.
In particular, THz hardware such as phase shifters and RF-chains possess high manufacturing costs and energy consumption.
Inspired by this, the basic component of THz ULAA could convert to subarrays, instead of antennas as in the microwave band, typical structures including array-of-subarray (AoSA), dynamic array-of-subarray (DAoSA), WSMS and dynamic-subarray with fixed-true-time-delay (DS-FTTD)~\cite{ref_hybrid_beamforming}.
Each RF-chain could dynamically connect to one or several subarrays through switches and phase shifters, such as DAoSA and DS-FTTD.
The subarray-based ULAA design reduces the required number of phase shifters and RF-chains, leading to low hardware complexity and power consumption.
From THz wave propagation perspective, compared to traditional structures without the subarray units, the beamforming gain and transmission flexibility are guaranteed in these subarray-inspired architectures through dedicated beamforming algorithm design~\cite{ref_hybrid_beamforming}. Specifically, the disjoint subarrays possess capabilities to independently generate narrow beams with high beamforming gain, helping overcome the propagation loss.
Moreover, subarray coordination can be simultaneously conducted through digital processing, which provides flexibility in transmission design.

With such ULAA architecture, several structural constraints impose challenges to cross-field communication design.
As illustrated in Fig.~\ref{fig_capacity_comp}, the channel capacity increases in the near-field with the decrement of distance, mainly due to the additional spatial degree-of-freedom (SDoF) brought by the non-linear channel phases~\cite{ref_hybrid_beamforming}.
The SDoF enhances the multiplexing capability and thus the channel capacity.
However, to unleash this near-field spatial multiplexing potential, more RF-chains are needed compared to the far-field scenario, which causes inconsistent design requirements, i.e., low hardware complexity and power consumption of the THz ULAA architecture.
Moreover, as only a small number of RF-chains are deployed compared to the number of antennas, 
the received signals are severely compressed from the antenna dimension to the RF-chain dimension.
To obtain complete channel information, observations of the antenna dimension are required, which, however, results in overwhelmed pilot overhead for channel estimation, as multiple pilot transmissions comparable to the number of antennas are needed.
In addition, beam management is jointly conducted by analog beams generated by subarrays and digital baseband.
In cross-field communications, both hardware constraints, e.g., constant module constraint of the phase shifter, and far- and near-field effects including angle and distance resolutions, need to be carefully addressed.

\section{Cross-field Solutions in THz ULAA Systems}
\label{sec_hybrid_solutions}

In this section, we introduce three cross-field technologies based on subarray architecture in THz ULAA systems, including the HSPM channel model, CS-based cross-field channel estimation and WSMS hybrid beamforming.

\subsection{HSPM Channel Model}
Channel models act as the fundamental of transmission design.
To assist THz ULAA cross-field communications, a channel model should balance high accuracy and low complexity, and be efficiently suitable for far- and near-field.
In THz ULAA systems, the subarrays are regarded as basic units.
A physical subarray usually contains substantially fewer antennas than the entire ULAA, e.g., 64 elements.
As a result, the Rayleigh distance for a subarray is typically small, e.g., 0.05m for a 64-element square shape subarray operating at 0.3~THz.

\begin{figure}[t]
	\centering
	{\includegraphics[width= 0.48\textwidth]{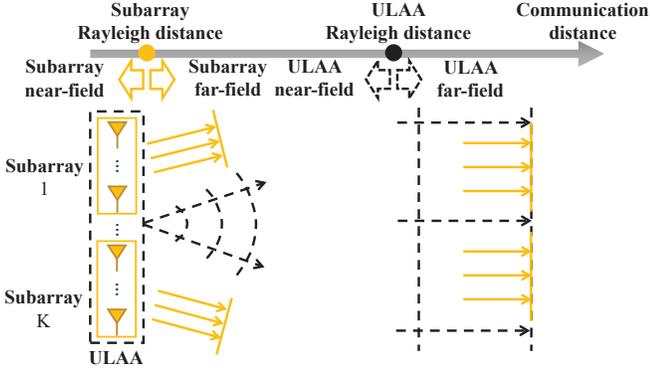}}
	\caption{An illustration of HSPM channel model.} 
	\label{fig_TLAA_Rayleigh}
	\vspace{-5mm}
\end{figure}

Inspired by these facts, an HSPM is proposed in~\cite{ref_HSPM}, as shown in Fig.~\ref{fig_TLAA_Rayleigh}.
HSPM employs PWM within one subarray.
The PWM-based modeling remains precise since it is reasonable to consider that transmissions are always conducted in the far-field originating from a subarray.
By contrast, SWM is deployed among subarrays to address the near-field spherical-wave propagation for improved modeling accuracy, since communications are conducted in a cross far- and near-field of the entire ULAA. 
Due to the deployment of hybrid PWM and SWM modeling, HSPM is accurate compared to PWM in both far- and near-field regions.

In terms of modeling complexity, 
the required number of parameters to determine HSPM is proportional to the production of subarrays at Tx, Rx, and the number of propagation paths in the channel, which is much smaller than that of the SWM.
Although the modeling complexity of HSPM is slightly higher than PWM, by further dividing the physical subarray into several virtual subarrays, deploying PWM within the virtual subarray, and SWM among virtual subarrays during the modeling process, HSPM is effective in achieving balanced accuracy and complexity. 

\subsection{Cross-field Channel Estimation}
It is necessary to obtain accurate channel state information at the antenna level, which helps to determine the analog and digital precoding.
To observe a high dimensional channel with a small number of RF-chains, multiple pilot transmissions are needed, resulting in high pilot overhead.
To solve this, CS-based channel estimation methods were widely explored in both far- and near-field communications~\cite{ref_CE_ELMaMIMO,ref_OMP}.
By exploiting the sparsity of the THz ULAA wireless channel, these schemes can recover the high-dimensional channel matrix from compressed observation and achieve low pilot overhead.

To deploy the CS-based methods, the wireless channel needs to be sparsely represented by a codebook. 
In near-field transmission with SWM, both angular and distance dimensions are sampled, and each sampled point relates to one grid, i.e., one realization of the codebook.
The sparse representation codebook is constructed by collecting the array steering vectors of these points.
As the number of grids is usually much larger than the propagation paths, the channel could be sparsely represented.
However, this two-dimensional sampling leads to excessive codebook dimension, which brings overwhelming operational complexity~\cite{ref_CE_ELMaMIMO}.
By contrast, for far-field PWM, a path's incident angle is considered the same for the entire ULAA. Therefore, one-dimensional sampling on the angular domain is enough to compose the codebook~\cite{ref_OMP}.
Nevertheless, this sampling strategy mismatches the near-field environment, resulting in low estimation accuracy in the cross-field propagation environment.

The subarray units in ULAA open new access to CS-based cross-field channel estimation~\cite{ref_DSE_SSE, ref_Hybrid_CE_GC}.
Specifically, within each subarray, the incident angle of a path is considered to be the same for all subarray elements. 
Therefore, during the operation, the angular domain of the subarray channel is sampled to compose a subarray codebook.
Moreover, among subarrays, the incident angle of a path is considered to be different for different subarrays.
To account for this, the angular samplings for different subarrays are independent. 
In this case, both near- and far-field channels can be sparsely represented by the subarray-based method.
The proposed sparse representation method possesses the same complexity as traditional solutions based on PWM, which, however, owns much higher accuracy in the near-field.
Moreover, compared to the joint angular and distance domain sparse representation using SWM, the computational complexity of the proposed codebook is significantly reduced. 

After that, different channel estimation algorithms could be designed to recover the antenna level channel.
For example, the separate side estimation (SSE) in~\cite{ref_DSE_SSE} decouples the non-zero grid estimation at Tx and Rx to achieve low complexity. 
Moreover, the dictionary shrinkage estimation (DSE) in~\cite{ref_DSE_SSE} exploits the spatial correlation among subarrays, i.e., the incident angles for different subarrays are close in the spatial domain, to further reduce the complexity.
In addition, the fixed point iteration, as well as deep learning tools, are deployed in~\cite{ref_Hybrid_CE_GC} to estimate the cross-field
channel with high efficiency. 
Notably, in wideband systems, the channel response varies with carrier frequencies.
The wideband channel estimation can be conducted by dividing multiple subcarriers. In this case, the sub-channel on each subcarrier could be regarded as invariant and estimated by the above channel estimation methods.

\subsection{WSMS Hybrid Beamforming}
\label{subsec_WSMS_hybrid_beamforming}
With strong sparsity in the THz channel, the spatial multiplexing capability in the far-field is upper bounded by the number of propagation paths, i.e., the \emph{inter-path multiplexing}, instead of the number of antennas as in the microwave band. 
However, it is observed in Fig.~\ref{fig_capacity_comp} that the channel capacity gap between SWM and PWM increases by reducing the communication distance.
This is caused by the additional spatial multiplexing gain brought by spherical-wave propagation in the near-field region, as explained in Sec.~\ref{subsec_challenge_architecture}, namely, the \emph{intra-path multiplexing}.
Specifically, when spherical-wave propagations are introduced, a notable effect is that the phases of columns/rows in the channel matrix for a particular propagation path become linearly independent~\cite{ref_hybrid_beamforming}.
This results in additional SDoF thus multiplexing gain that can be exploited for one propagation path among different antennas. Since the intra-path multiplexing gain is not limited by the number of multipath, it is naturally applicable and promising for THz communications with a notable channel sparsity. Therefore, transmissions in the near-field by processing higher spectral efficiency with both inter- and intra-path multiplexing are preferred for THz ULAA systems~\cite{ref_hybrid_beamforming}.

However, the near-field range is limited, only covering a certain proximal range, especially in the outdoor scenario. For example, within 1m by equipping compact UPA with 1024 elements at Tx and Rx in Fig.~\ref{fig_capacity_comp}. 
To further enhance the spatial multiplexing even in the far-field of the compact array, a WSMS hybrid beamforming architecture is studied in~\cite{ref_hybrid_beamforming}. 
Unlike traditional architectures where antenna spacing of the entire array equals half of the wavelength, the WSMS enlarges subarray spacing to tens or even hundreds of times half-wavelength. The antenna spacing within the subarray is set as half of the wavelength. 
In this way, the near-field region is expanded to tens to hundreds of meters. 
Compared to the far-field transmissions, the total multiplexing gain is improved by a factor equal to the number of subarrays.
In conclusion, by jointly exploiting the inter-path and intra-path multiplexing, this co-located or distributed WSMS can significantly enhance spatial multiplexing and, thus, data rate.

In terms of WSMS hardware manufacture, the RF-chains and phase shifters are equipped for each subarray.
Within the subarray, each RF-chain connects to several phase shifters, which number equals the antennas. 
Moreover, the antennas, phase shifters, and RF-chains are disjoint from each other. 
Therefore, the analog beamforming matrix holds a block-diagonal structure, with the phase of each element satisfying constant module constraint. Those constraints are addressed in the analog and digital beamforming algorithms design.
It is worth noticing that the WSMS operates with small fractional bandwidth, i.e., the ratio between the bandwidth and central frequency.
For example, at 5 GHz bandwidth at 0.3 THz carrier frequency.
In THz wideband systems with large fractional bandwidth, a kind of beam-squint effect that could reduce the array gain and spectral efficiency should be carefully addressed~\cite{ref_hybrid_beamforming}.

\section{Performance Evaluation}
\label{sec_performance_evaluation}

In this section, we evaluate the performances of HSPM, subarray-based channel estimation method and WSMS hybrid beamforming structure.
During our simulation, the planar-shaped THz ULAA is deployed at both Tx and Rx.
We choose a carrier frequency at 0.3~THz, and the number of multi-path is set as 2~\cite{ref_hybrid_beamforming}.
In the WSMS structure, the number of subarrays and RF-chains is set as 4, and the number of antennas in ULAA is 1024 unless otherwise specified. Therefore, the number of antennas of a subarray equals 256.
Moreover, the bandwidth is selected as 5GHz.

\begin{figure}[t]
	\centering
	{\includegraphics[width= 0.4\textwidth]{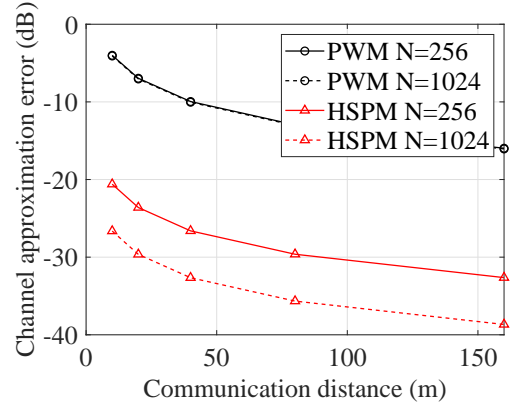}}
	\caption{Channel approximation error in different communication distances. Subarray spacing is fixed at 32 times wavelength.} 
	\label{fig_channel_error}
	\vspace{-5mm}
\end{figure}

To illustrate the advantage of HSPM in ULAA systems, we numerically evaluate the approximation errors of HSPM and PWM in different communication distances by calculating their differences from the ground-truth SWM.
The approximation error is obtained as 
$\Vert \mathbf{H}_{\rm M} - \mathbf{H}_{\rm S} \Vert_{\rm F}^2/\Vert \mathbf{H}_{\rm S} \Vert_{\rm F}^2$,
where $\mathbf{H}_{\rm M}$ and $\mathbf{H}_{\rm S}$ denote the adopted channel matrix based on HSPM or PWM and the SWM channel matrix, respectively, $\Vert \cdot \Vert_{\rm F}$ depicts the Frobenius norm.
Notably, the approximation error of the SWM equals 0 in this case.
As shown in Fig.~\ref{fig_channel_error},
HSPM possesses much higher modeling accuracy compared to PWM.
With 1024 antennas in the system, the approximation error of HSPM is 22.5dB lower than PWM at 40m.
Moreover, we calculate the number of parameters required for representing different channel models.
In the considered setup with 1024 antennas and 4 subarrays at BS and UE, the required number of parameters for SWM and PMW are around $4\times 10^6$ and $12$, respectively. 
For comparison, the required numbers of parameters for HSPM is around $400$.
Therefore, we can state that HSPM achieves around 175 times higher accuracy than PWM, while it deploys $0.01 \%$ and 33 times the number of parameters of SWM and PWM, respectively. 

\begin{figure}[t]
	\centering
	{\includegraphics[width= 0.4\textwidth]{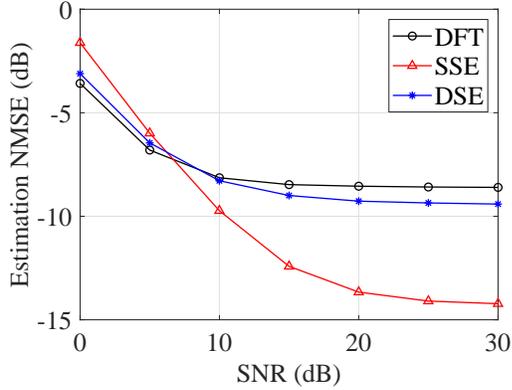}}
	\caption{Comparision of different channel estimation methods.}
	\label{fig_channel_est_comp}
	\vspace{-5mm}
\end{figure}

In Fig.~\ref{fig_channel_est_comp}, the channel estimation normalized mean-square-error (NMSE) based on traditional discrete Fourier transform (DFT)~\cite{ref_OMP}, proposed SSE and low complexity DSE~\cite{ref_DSE_SSE} are evaluated, respectively.
We observe that channel estimation based on the proposed subarray codebook including SSE and DSE possesses much higher estimation accuracy than the traditional method. 
At SNR=15dB, the SSE possesses 4dB lower NMSE than the traditional DFT-based method, validating its superiority.
Moreover, the proposed DSE also achieves 1dB lower NMSE than the traditional method.
Complexities of the far-field, subarray-based, and low complexity methods are obtained as $\mathcal{C}(N^2)$, $\mathcal{C}(N)$ and $\mathcal{C}\left( \frac{N}{K}\right)$, respectively, where $N$ and $K$ denote the number of antennas and subarrays in the THz ULAA system.
Notably, both SSE and DSE are based on on-grid sparse channel representation codebooks that take the subarray as a unit. The estimation accuracy of which is strictly limited by the grid resolution. 
Although increasing the number of grids or performing grid refinement could further improve the estimation accuracy, it comes at the cost of higher computational complexity, which is not preferred in THz UM-MIMO systems with a massive number of antennas~\cite{ref_DSE_SSE}.

\begin{figure}[t]
	\centering
	{\includegraphics[width= 0.4\textwidth]{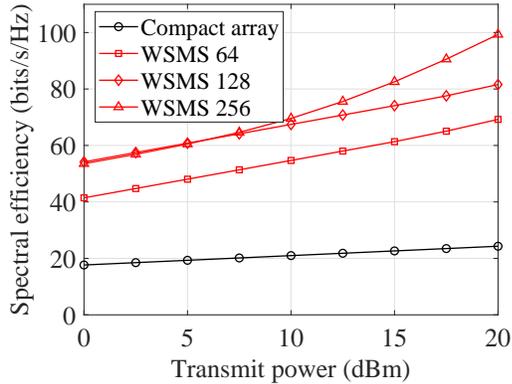}}
	\caption{Spectral efficiency in different hybrid beamforming architectures.}
	\label{fig_SE_WSMS_FC}
	\vspace{-5mm}
\end{figure}

Finally, in Fig.~\ref{fig_SE_WSMS_FC}, we compare the spectral efficiency between WSMS and traditional compact array systems in the literature.
It is worth noticing that the compact array refers to the array with half-wavelength antenna spacing. Most existing array structures belong to this category, e.g., the fully connected (FC), array-of-subarray (AoSA) and DAoSA~\cite{ref_hybrid_beamforming}.
4 widely-spaced subarrays are deployed in the WSMS architecture. The communication distance is fixed at 40m.  
The spectral efficiency is obtained by calculating channel capacities based on different structures, which values reflect the upper bounds of spectral efficiency.
In Fig.~\ref{fig_SE_WSMS_FC}, the subarray spacing is set as 64, 128 and 256 times of wavelength, respectively. It is observed the multiplexing gain is enhanced by the widely spaced architectural design.
Specifically, the spectral efficiency of the THz WSMS hybrid beamforming architecture is 3.2 times higher than the traditional compact array counterpart at 15 dBm transmit power and $128\lambda$ subarray spacing.
It is further observed that by considering HSPM, the spectral efficiency increases with the WSMS subarray spacing. This is due to the fact that a larger subarray spacing leads to greater differences among the sub-channels of subarrays, which helps enhance the multiplexing gain.

\section{Open Problems and Potential Research Directions}
\label{sec_open_problems}
In this section, we elaborate open problems and future research directions for THz ULAA cross-field communications.

\subsection{Theoretical Bounds}
Existing studies have experimentally demonstrated that ULAA hybrid-field communications have the potential of achieving performance enhancement, including channel modeling accuracy, channel estimation NMSE and spectral efficiency~\cite{ref_HSPM,ref_DSE_SSE,ref_hybrid_beamforming,ref_Hybrid_CE_GC}. 
However, as fundamental to communication system design, the theoretical bounds of the cross-field technologies, such as system capacity, degree of freedom, estimation Cramer-Rao lower bound (CRLB), etc., are still lacking.
For example, it is worth exploring the CRLB and proposing effective cross-field parameter estimation algorithms to approach it. Moreover, the training and system configuration could be designed with the aim of decreasing the CRLB.
In all previous analyses, the effects of subarray division, subarray distance, and communication distance need to be explored and possibly optimized to guide the system design.
In addition, it is worth systematically analyzing and evaluating the overall cost and benefit of the subarray-based solution in the cross-field by jointly taking the hardware limitation, power consumption, and capacity enhancement into consideration.

\subsection{Practical Hardware Efficient Design}
The hardware difficulty and cost for THz ULAA hardware grow simultaneously as communication frequency moves to the THz band.
This inspires the use of fewer RF-chain and phase shifters to compose the hybrid ULAA architecture~\cite{ref_25_Heath}.
However, as analyzed in Sec.~\ref{subsec_challenge_architecture}, more RF-chains and phase shifters are needed to unleash the spatial multiplexing capability in the near-field,
which contradicts the low hardware cost requirement.
In the meantime, the required number of RF-chains in the far-field is smaller compared to the near-field case, due to the limited SDoF.
As the ULAA hardware structure is usually fixed, it is difficult to simultaneously meet the cross-field spatial multiplexing and hardware cost requirements.
As two potational solutions, the WSMS introduced in Sec.~\ref{subsec_WSMS_hybrid_beamforming} enlarges the near-field region to the typical communication distance. 
Moreover, a distance-aware precoding structure (DAP) was proposed~\cite{ref_nearfield_dai}.
By inserting a selection circuit between phase shifters and RF-chains, each RF-chain can be flexibly activated or inactivated according to the SDoFs.
Although both structures could be deployed for cross-field transmissions, the required number of RF-chains and phase shifters are larger than the structures designed for far-field transmission.
It is still worth exploring a more efficient structure to balance spatial multiplexing and hardware cost in cross-field transmissions.

\subsection{Cross-field Beam Training}
To establish reliable ULAA transmissions,
a beam-searching process, known as beam training, is required to realize narrow beam alignment.
{In the far-field, the beams only process angular resolution and beam training searches the angular domain to find the best beam pair~\cite{ref_subarray_lin}.
To perform near-field beam training, joint searchings in the angular and distance domains are conducted in the literature~\cite{ref_nearfield_dai,ref_near_training}.
On one hand, as the searching beams are obtained from a pre-defined beam codebook, the far-field codebook designs are unable to meet the joint angular and distance resolutions in the cross-field. On the other hand, although the near-field codebooks can be deployed for far-field, the additional distance domain searching undoubtedly increases the training overhead despite the codebook design and training strategy,  which incurs high complexity and is thus not preferred in cross-field.}
To study the efficient unified cross-field beam training framework, first, the angular and distance domain beam patterns in the cross-field should be identified, by considering different communication distances and ULAA structure.
Second, {to achieve low training overhead, it is worth exploring the suitable training beams for alignment. For example, the study of hierarchical codebooks with non-uniformly adjustable beamwidth in angular and distance domains~\cite{ref_nearfield_dai}, or directly using the far-field beam for training and proposing beam estimation algorithms for alignment.}
Moreover, the training strategy for both single and multi-user scenarios, as well as beam tracking strategies under user movement needs further study.

\subsection{Multi-user Communications and Networking}
Although THz cross-field communication has been studied for physical layer single link transmission~\cite{ref_hybrid_beamforming}, its impact on medium access control (MAC) and networking has not drawn enough research attention.
A recent study on multiple access {in the near-field}~\cite{ref_Dai_LDMA} has shown that compared to the spatial division multiple access (SDMA) in the far-field, the multiple accessibilities could be enhanced by the new distance dimension division ability in the near-field, which is named as location division multiple access (LDMA).
However, as the communication distance crosses near- and far-field, it is necessary to study the unified cross-field multiple access, by accounting for distance and angular beam resolutions in different communication distances.
For example, dividing the use of the LDMA and SDMA with a distance threshold.
Moreover, due to the change of physical layer characteristics, in the networking process, universal schemes with angular-distance-multiplxing are needed, to meet the transmission coverage, and deafness avoidance as well as control channel selection in the node discovery and coupling process.
The cross-field impacts on interference cancellation, network topology as well as routing require further study.

\section{Conclusion}
\label{sec_conclustion}

THz ULAA systems are widely exploited by effectively combating the distance limitation and blockage problem.
With the enlarged dimension of ULAA, typical communications distances reach both far- and near-field, resulting in a new paradigm of cross-field communications.
In this article, we investigate the challenges and features of ULAA hybrid-field communications from channel and ULAA architecture perspectives. 
The cross-field HSPM, CS-based channel estimation, and WSMS hybrid beamforming framework are introduced and evaluated. 
Finally, multiple open problems and potential research directions to ULAA cross-field communications awaiting to be explored are presented, including theoretical bounds, hardware-efficient designs, cross-field beam training and multi-user communications and networking. 

\section{Acknowledgement}
This work is supported by the National Natural Science Foundation of China under Grants 62171280 and by
the Fundamental Research Funds for the Central Universities of China.

\section*{Biographies}
\vspace{-1cm}
\begin{IEEEbiographynophoto}{Chong Han}
    is John Wu \& Jane Sun Endowed Associate Professor with the Terahertz Wireless Communications (TWC) Laboratory
	and with Department of Electronic Engineering and Cooperative Medianet
	Innovation Center (CMIC), Shanghai Jiao Tong University, China.
\end{IEEEbiographynophoto}
\vspace{-8mm}
\begin{IEEEbiographynophoto}{Yuhang Chen}
is currently pursuing a Ph.D. degree in the Terahertz Wireless Communications (TWC) Laboratory, Shanghai Jiao Tong University, China.
\end{IEEEbiographynophoto}
\vspace{-8mm}
\begin{IEEEbiographynophoto}{Longfei Yan}
is currently pursuing a Ph.D. degree in the Terahertz Wireless Communications (TWC) Laboratory, Shanghai Jiao Tong University, China.
\end{IEEEbiographynophoto}
\vspace{-8mm}
\begin{IEEEbiographynophoto}{Zhi Chen}
is a Professor with University of Electronic Science and Technology of China.
\end{IEEEbiographynophoto}
\vspace{-8mm}
\begin{IEEEbiographynophoto}{Linglong Dai}
is an associate professor at Tsinghua University. He has received six conference best paper awards and four journal best paper awards.
\end{IEEEbiographynophoto}
\end{document}